\documentclass[10pt,twocolumn,nofootinbib,groupedaddress,showkeys,preprintnumbers,floatfix]{revtex4-2}

\usepackage{epsfig}
\usepackage{amsmath,amssymb}
\usepackage{multirow}
\usepackage{graphicx}
\usepackage[usenames,dvipsnames,svgnames,table]{xcolor} 
\newcommand{\nn}{\nonumber}
\newcommand{\fig}[1]{Fig.\,\ref{#1}}
\newcommand{\tabl}[1]{Table\,\ref{#1}}
\newcommand{\eqn}[1]{Eq.\,\eqref{#1}}
\newcommand{\refsec}[1]{Sect.\,\ref{#1}}
\newcommand{\ud}{\mathrm{d}}
\newcommand{\unit}[1]{\,\mathrm{#1}}

\newcommand{\nuc}[2]{\ifmmode{{}^{#2}\mathrm{#1}}\else{${}^{#2}\mathrm{#1}$}\fi}
\begin{document}

%--------------------------------------------------------------------------------------------------------------------------
%        Title 
%--------------------------------------------------------------------------------------------------------------------------
\title{Big Bang nucleosynthesis in a weakly non-ideal plasma}

%--------------------------------------------------------------------------------------------------------------------------
%       List of Authors
%--------------------------------------------------------------------------------------------------------------------------
\author{Dukjae Jang}
%\email{djjang2@ibs.re.kr}
\affiliation{Center for Relativistic Laser Science, Institute for Basic Science (IBS), Gwangju 61005, Republic of Korea}

%------------------------------------
\author{Youngshin Kwon}
\email{Corresponding author: {ykwon@kau.ac.kr}}
\affiliation{Research Institute of Basic Science, Korea Aerospace University, Goyang 10540, Republic of Korea \\
Center for Extreme Nuclear Matters (CENuM), Korea University, Seoul 02841, Republic of Korea}

%------------------------------------
\author{Kyujin Kwak}
%\email{kkwak@unist.ac.kr}
\affiliation{Department of Physics, Ulsan National Institute of Science and Technology (UNIST), Ulsan 44919, Republic of Korea}

%------------------------------------
\author{Myung-Ki Cheoun}
%\email{cheoun@ssu.ac.kr}
\affiliation{Department of Physics and OMEG Institute, Soongsil University, Seoul 06978, Republic of Korea}

%--------------------------------------------------------------------------------------------------------------------------
%        Date
%--------------------------------------------------------------------------------------------------------------------------
\date{\today}

%--------------------------------------------------------------------------------------------------------------------------
%        Abstract 
%--------------------------------------------------------------------------------------------------------------------------
\begin{abstract}
We propose a correction of the standard Big Bang nucleosynthesis (BBN) scenario to resolve the primordial lithium problem by considering a possibility that the primordial plasma can deviate from the ideal state. In the standard BBN, the primordial plasma is assumed to be ideal, with particles and photons satisfying the Maxwell-Boltzmann and Planck distribution, respectively. We suggest that this assumption of the primordial plasma being ideal might oversimplify the early Universe and cause the lithium problem. We find that deviation of photon distribution from the Planck distribution, which is parameterised with the help of Tsallis statistics, can resolve the primordial lithium problem when the particle distributions of the primordial plasma still follow the Maxwell-Boltzmann distribution. We discuss how the primordial plasma can be weakly non-ideal in this specific fashion and its effects on the cosmic evolution.
\end{abstract}

%--------------------------------------------------------------------------------------------------------------------------
%        Keywords 
%--------------------------------------------------------------------------------------------------------------------------

\maketitle

%--------------------------------------------------------------------------------------------------------------------------
%       Body Text 
%--------------------------------------------------------------------------------------------------------------------------

%=====================
\section{Introduction}
\label{s01}
%=====================
Big Bang nucleosynthesis (BBN) is one of the most convincing pieces of evidence for the hot Big Bang cosmology. The precise measurement of the cosmic microwave background (CMB) and the accordingly determined value of the baryon-to-photon ratio, $\eta = (6.094 \pm 0.063)\times 10^{-10}$~\cite{Ade:2015xua}, imply that there is no free parameter left in the standard model of BBN (SBBN), which describes the formation of the lightest nuclides, such as \nuc{D}{}, \nuc{He}{3}, \nuc{He}{4}, and \nuc{Li}{7}.  The predicted abundances of SBBN are in remarkable agreement with the spectroscopic observations for deuterium and helium isotopes. Thanks to this success of SBBN, BBN plays a crucial role in constraining new physics beyond the standard model of particle physics and cosmology~\cite{Cyburt:2004yc}, such as dark matter searches, dark energy searches, and so on.%

However, there remains an unsolved problem: the relic abundance of \nuc{Li}{7} inferred from observations is significantly below the predicted value of SBBN. This discrepancy, referred to as the primordial lithium problem, could reflect a difficulty in observationally determining the primordial lithium abundance~\cite{Howk:2012rb}, or make room for new physics providing a modification of the SBBN~\cite{Kusakabe:2014moa}. While these possible solutions involving additional hypotheses still need to be cross-verified experimentally or observationally, it is worth checking if the basic assumptions present in the SBBN were oversimplifying the early Universe and thus overestimating the lithium abundance.%

In the current scenario of the SBBN~\cite{Cyburt:2015mya}, the formation of the light elements is initiated by the neutrino freeze-out during the radiation-dominated era  and lasts until the temperature of the primordial plasma is $\sim 10^{7}\unit{K}$. (This temperature corresponds to a timescale of a few minutes in the cosmic evolution.) At this epoch, the Universe is assumed to evolve with an idealised thermodynamic process. The assumption of the adiabatic expansion of the Universe without any heat flux into it is reasonable. During the expansion, the Universe continues to find a new equilibrium as long as the thermonuclear reaction rates are greater than the cosmic expansion rate. Such an assumption of the isentropic expansion of the Universe, which is ideal, seemingly stands to reason. However, taking idealisation a step further should be done more carefully. For instance, constituents of the primordial plasma are usually treated as ideal gases in the SBBN. However, this assumption of ideal plasma could be questionable.%

\begin{figure}[t]
	\centering
		\includegraphics[width=8cm]{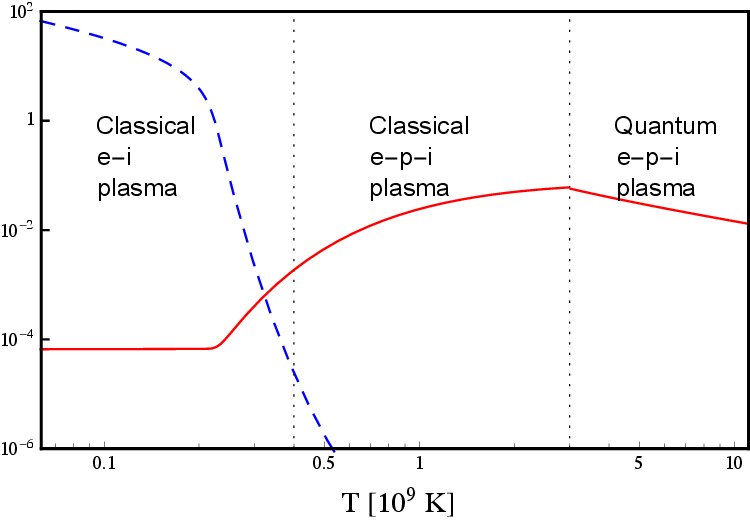}
		\caption{Plasma parameter (red solid line) and electron chemical potential in unit of $\mu^{}_e/(k^{}_B T)$ (blue dashed line) during the BBN epoch. The plasma constituents are indicated by e-i (electron-ion) and e-p-i (electron-positron-ion) for respective temperature regions.}
		\label{fig1}
	\end{figure}

A classical plasma can be characterised by the plasma parameter, $\Gamma=n^{1/3}_e e^2/(k^{}_BT)$, which is defined as the ratio of the mean potential energy to the thermal kinetic  energy~\cite{Gibbon:2016qcp}, where $n^{}_e$, $e$, $k^{}_B$, and $T$ stand for the free electron density, elementary electric charge, Boltzmann constant, and temperature, respectively. The plasma of $\Gamma\gtrsim 1$ is classified as a non-ideal plasma, whereas the ideal plasma corresponds to the case of $\Gamma \ll 1$. Figure\,\ref{fig1} shows the plasma parameter as a function of temperature during the BBN epoch. At high temperatures  ($T\gtrsim 5\times10^9\unit{K}$), higher than the mass scale of a free electron ($0.5\unit{MeV}$), the primordial plasma is a quantum plasma that contains degenerate electrons for which the plasma parameter is rewritten by replacing the thermal kinetic energy with the Fermi energy, $\varepsilon^{}_F \propto n^{2/3}_e$~\cite{Fortov:2000}. Then, the plasma parameter behaves reciprocally to the classical plasma, increasing with decreasing density, $\Gamma=n^{1/3}_e e^2/ \varepsilon^{}_F \propto n^{-1/3}_e$. At the temperatures lower than the electron mass scale, the plasma can be treated classically. Since the free electron density decreases rapidly with the evolution of the Universe, the plasma is said to be ideal in the low-temperature BBN era ($T\lesssim 10^8\unit{K}$). The plasma in the moderate-temperature region would be regarded as weakly non-ideal with the plasma parameter of the order of $10^{-2}$.%

Opher and Opher~\cite{Opher:1997mu} pointed out that the collective effects, such as plasma waves, in the weakly non-ideal plasma of $T\geq 10^{9}\unit{K}$ could invoke meaningful deviations from an ideal blackbody curve at low frequencies of the electromagnetic spectrum. They selected two different plasma states linked to different temperature regions in the BBN epoch: the electron-positron plasma at $T=10^{10}\,\unit{K}$ and the electron-proton plasma at $T=10^9\unit{K}$, respectively. One of their findings is that the distortion of the photon spectrum at low frequencies is more severe in the electron-proton plasma than in the electron-positron plasma. Leaving aside the argument that, as indicated in \fig{fig1}, plenty of positrons could still survive at $T=10^9\unit{K}$, it is worth mentioning that their results originate in the different collective effects of positrons and ions. As explored in the laboratory and space plasma, heavy ions including protons respond more slowly to perturbations than electrons do, and a relatively slow response of the heavy ions can lead to the self-generated electric fields and/or the ion acoustic waves which could intensify the wave-particle interaction. In contrast, positrons respond to any perturbations as instantly as electrons do, mitigate collective effects, and stabilise the plasma relatively quickly.%

In addition to the plasma parameter, a non-ideal state of the BBN plasma can also be shown with the chemical potentials of the plasma constituents. In \fig{fig1}, we show how the electron chemical potential changes during the BBN epoch. The change is more dramatic towards the lower temperature region ($T \lesssim 10^9\unit{K}$). We should note that there must be indispensable instability at $T\lesssim 10^9\unit{K}$, which is caused by a chemical disequilibrium between the Dirac ($e^+e^-\to2\gamma$) and the Breit-Wheeler ($2\gamma\to e^+e^-$) processes and lasts a while until the positrons are annihilated in the early Universe. The chemical potential of electrons is known to become large, $\mu^{}_e/(k^{}_B T) \gtrsim 1$, at $T\lesssim 10^8\unit{K}$ due to the charge neutrality and positron annihilation at low temperatures~\cite{Pitrou:2018cgg}. Such changes in fugacity indicate that the primordial plasma of degenerated electrons at $T \lesssim  10^9\unit{K}$ possibly deviates from ideal plasma. In such a non-ideal plasma caused by the severe fugacity changes, the photon spectrum can also be distorted and calculated (or traced as a function of time) by directly solving the Boltzmann equation with the time-dependent source term of pair annihilation included. Such a detailed study is not available yet, but the non-ideal state of the BBN plasma could possibly appear at the late BBN epoch ($T \lesssim 10^9\unit{K}$), during which light elements were still actively synthesised.%

The problem we face in \fig{fig1} is that the plasma parameter and the electron chemical potential are mutually incompatible. For example, the plasma parameter at $T\sim10^8\unit{K}$ was evaluated for a classical plasma of $\varepsilon^{}_F /(k^{}_B T ) \ll 1$. The electron chemical potential curve, on the other hand, shows that $\mu^{}_e/(k^{}_B T) \geq 1$ at $T\sim10^8\unit{K}$. However, the consequent inequality $\varepsilon^{}_F \ll \mu^{}_e$ is not self-consistent. This implies that the BBN plasma might be more complicated than in the assumption of the SBBN, and establishing when the non-ideal plasma occurs in the BBN era requires in-depth investigation, which is beyond the scope of the current work.%

By considering the above possibility, in this study, we investigated whether the non-ideal state of the late BBN plasma can resolve the \nuc{Li}{7} problem. Specifically, we assumed that the non-ideality distorted the electromagnetic spectrum at high frequencies only. Again, the exact behaviour of the photon spectrum at $T \lesssim 10^9\unit{K}$ must be obtained by solving the Boltzmann equation, which is beyond the scope of the current work. We note that the distortion of the photon spectrum at low frequencies has only a small effect on the abundances of light elements synthesised during the BBN epoch simply because the photo-disintegration rates (the most dominant effect that the distorted photon spectrum could have on the SBBN prediction) can only change meaningfully when the photon spectrum deviates from the Planck distribution at high frequencies. However, the distorted photon spectrum, at both low and high frequencies, affects the total energy density of photons in the early Universe, which influences the evolution of the early Universe as well.%

The rest of this paper is organised as follows. In \refsec{s02}, we propose a model that could resolve the primordial lithium problem with the non-ideal plasma state of the late BBN epoch. The results (i.e. the primordial abundances) obtained with our model are presented in \refsec{s03}. In order to be consistent with the standard cosmology, particularly the observational CMB, it is necessary to restore the distorted photon spectrum to the Planck distribution, which is discussed in \refsec{s04}. Finally, conclusions are drawn in \refsec{s05}.%

%=====================
\section{Model}
\label{s02}
%=====================
Out of twelve main reactions in the SBBN~\cite{Cyburt:2015mya}, only four are affected by a photon distribution in their inverse reactions: 
	\begin{eqnarray}
		\mathrm{p} + \mathrm{n} &\to& \mathrm{D} + \gamma~, \label{eq1}\\
		\mathrm{D} + \mathrm{p} &\to& {}^{3}\mathrm{He} + \gamma~, \label{eq2}\\
		{}^{3}\mathrm{He} + {}^{4}\mathrm{He} &\to& {}^{7}\mathrm{Be} + \gamma~, \label{eq3}\\
		\mathrm{T} + {}^{4}\mathrm{He} &\to& {}^{7}\mathrm{Li} + \gamma~. \label{eq4} 
	\end{eqnarray}
It is noticeable that the sequence of the nuclear reactions, \eqref{eq1}-\eqref{eq3}, is in order to produce \nuc{Be}{7}. The theoretical excess in the final abundance of \nuc{Li}{7} is mainly attributable to the radioactive decay of the entire \nuc{Be}{7} to \nuc{Li}{7} via electron capture. If the photon distribution in the BBN plasma is slightly distorted this way in order to obtain a little extra energy due to the collective modes of the plasma (which is able to enhance the rates of photo-disintegration reactions), the enhanced rates will reduce the abundance of \nuc{Be}{7} and thus of \nuc{Li}{7} eventually.

In this respect, we propose the BBN that undergoes a transition from ideal to weakly non-ideal plasma at some temperature, $T^{}_\text{tr}$ as a proper solution to the primordial lithium problem. Relatively hot BBN at $T>T^{}_\text{tr}$ can be treated as in the SBBN. The plasma non-ideality is unveiled in the cool BBN at $T<T^{}_\text{tr}$.

In order to represent the plasma non-ideality below $T^{}_\text{tr}$, we distorted the photon distribution from the Planck distribution, adopting a generalised Planck distribution by Tsallis statistics~\cite{Tsallis:1987eu,Tirnakl:1997}:
	\begin{equation}
		f^{}_q = \frac{1}{\left[ 1-(1-q)\frac{E}{k^{}_BT} \right]^{\frac{1}{q-1}} -1}~,
		\label{eq5}
	\end{equation}
where $q$ determines the extent of the deviation. For example, the value of $q>1$ causes enhancement in the high-energy tail of the distribution function, while the standard Planck distribution is easily recovered in the  limit of $q\to1$. Tsallis statistics, which was invented by Tsallis~\cite{Tsallis:1987eu} and applied for the description of non-equilibrium statistical systems, has been used in the BBN calculation \cite{Bertulani:2012sv,Hou:2017uap,Kusakabe:2018dzx} to describe non-Maxwellian velocity distributions for matter particles in a non-thermal plasma. Here, we want to clearly emphasise that we did not change the standard Maxwell-Boltzmann (MB) distribution for the nuclei velocity distribution since the distribution function of the nuclei at thermodynamic equilibrium is given by the Maxwellian distribution~\cite{Pitrou:2018cgg}, and the deviation by thermalisation processes, such as electron-nucleus scattering, is inconsiderable~\cite{McDermott:2018uqm,Sasankan:2019oee}. 

In the BBN epoch, the plasma constituents are supposed to share a single plasma temperature as they are thermally coupled. This holds true even for $q\neq1$ in our calculation such that Coulomb collisions occur more rapidly than the cosmic expansion rate. Without loss of generality, the photon temperature defined in \eqn{eq5} is assumed to represent the common temperature for electrons and ions as well.

We chose Tsallis statistics in \eqn{eq5} because it provides convenience when modifying high-energy tails of photon distributions that we assumed for the non-ideal plasma. Furthermore, as mentioned earlier, since photons produced via positron annihilations are energetic compared to the average photon energy, the true photon spectrum in the non-ideal plasma in our consideration might be close to \eqn{eq5}. The standard Boltzmann-Gibbs statistics states that, if both kinetic and chemical equilibrium obtain, the solution of the Boltzmann equation for photons becomes the Planck distribution, that is, the Bose-Einstein distribution with vanishing chemical potential. Indeed, we are now concerned with the photons under kinetic equilibrium but falling out of chemical equilibrium due to the pair annihilation process. This can be signified with a frequency-dependent chemical potential in the Bose-Einstein statistics; lowering the rate of establishment of equilibrium for high frequencies. \eqn{eq5} at any rate is simple but good enough to test the distortion of high-frequency regions in the photon spectrum.

Now we assume a step function delineation in which $T^{}_\text{tr}$ separates the weakly non-ideal state ($T\le T^{}_\text{tr}$) from the ideal plasma ($T>T^{}_\text{tr}$):
	\begin{equation}
		q(T) =  \theta(T-T^{}_\text{tr}) + \theta(T^{}_\text{tr} - T)\,q\,'~,
		\label{eq6}
	\end{equation}
with $q\,'\neq1$. In reality, the transition from ideal to weakly non-ideal plasma could occur smoothly as the Universe cools down. Although the step function delineation between ideal and weakly non-ideal plasma seems schematic, it is valid as the simplest ansatz because it does represent the transition to the weakly non-ideal plasma state in BBN. We tested a ramping function instead of the step function to describe the smooth transition and found that the difference between them is not significant with regard to physical interpretation. The global temperature-dependence of $q$ is indeed endowed with temporal dependence. One also choose to consider the spatial dependence of $q$ to represent inhomogeneity or anisotropy. We first neglected higher order corrections such as the temperature gradient due to the anisotropy in the radiation field and assumed that the density contrast is not high enough to affect the final BBN abundances since a spatial fluctuation of $<20\unit{\%}$ in the photon number density would not change the order of the baryon-to-photon ratio. With this lead-up, the strategy pursued in this study is reduced to finding a satisfactory combination of ($q\,'$, $T^{}_\text{tr}$) for the BBN abundances.

The modified photon distribution, \eqn{eq5} with $q\neq1$, changes the following physical quantities: reaction rates of photo-disintegration processes, photon number, and energy density in comparison with the standard case of $q = 1$.

	\begin{figure}[t]
	\centering
		\includegraphics[width=8cm]{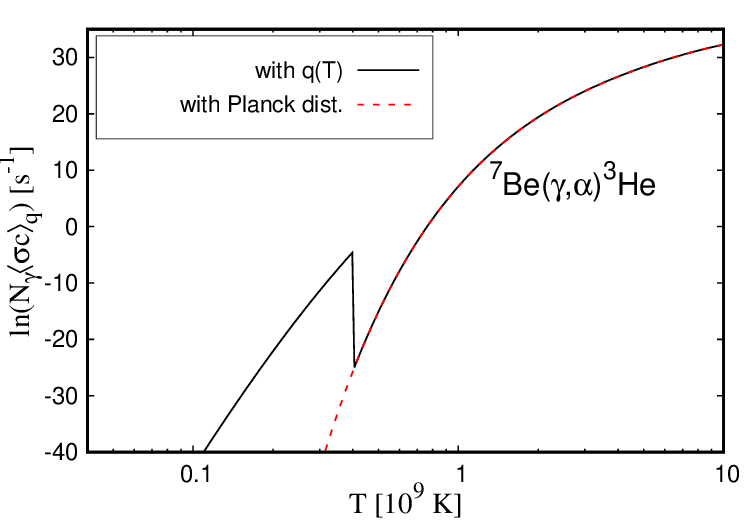}
		\caption{Reaction rate of $\nuc{Be}{7}(\gamma, \alpha)\nuc{He}{3}$ as a function of temperature. Red dashed line for $q=1$ corresponds to the standard Planck distribution and black solid line for $q\,'=1.027$ and $T^{}_\text{tr}=4\times10^8\unit{K}$.}
		\label{fig2}
	\end{figure}

	\begin{table}[b]
	\centering
	\begin{tabular}{|c|c|c|cc|}
		\hline
		~ & ~SBBN~ & ~This work~ & ~~~~~Observation~ & \\
		\hline
		~$Y_\text{p}$~							& ~$0.2474$	~	& ~$0.2474$	~	& ~$0.2446 \pm 0.0029$ & \cite{Peimbert:2016bdg} \\
		~D/H $(10^{-5})$~   					& ~$2.493$~	& ~$2.525$~	& ~$2.527 \pm 0.03$	& \cite{Cooke:2017cwo} \\
		~\nuc{He}{3}/H $(10^{-5})$	& ~$1.092$~	& ~$0.9253$	~	& ~$\leq 1.1 \pm 0.2$ & \cite{Bania:2002yj} \\
		~\nuc{Li}{7}/H $(10^{-10})$ 	& ~$5.030$~	& ~$1.677$~	& ~$1.58 \pm 0.31$	& \cite{Sbordone:2010zi} \\
		\hline
	\end{tabular}
	\caption{Calculated final abundances of \nuc{He}{4} ($Y_\text{p}$: mass fraction) and other nuclei (number ratio) relative to hydrogen in comparison with observational data, using the following input values: the baryon-to-photon ratio $\eta=6.031\times10^{-10}$,  neutron mean lifetime $\tau_n = 880.2\unit{s}$~\cite{Patrignani:2016xqp}, and effective number of neutrino species $N^{}_\text{eff}=3.046$~\cite{Mangano:2005cc}.}
	\label{tab1}
	\end{table}

For a reaction in the form of ($1+2\to 3+\gamma$) such as \eqref{eq1}-\eqref{eq4}, the reverse reaction rate with $q$, an example of which is displayed in \fig{fig2}, can be written as
	\begin{align}
		&N^{}_{\gamma q} \left\langle\sigma c\right\rangle^{}_{3 \gamma q} = \frac{m^{}_{12}}{\pi^2 \hbar^3} \,\frac{g^{}_1 g^{}_2}{g^{}_3(1+\delta^{}_{12})} \nn\\
		&\quad\times \int_{0}^{\infty} \sigma^{}_{12}(E) \,E \,\frac{1}{\left[ 1-(1-q)\frac{E+Q}{k^{}_BT} \right]^{\frac{1}{q-1}} -1} \,\ud E~,
		\label{eq7}
	\end{align}
where $m^{}_{12}$ is the reduced mass of particles $1$ and $2$, and $Q$ is the $Q$-value of the forward reaction. In \eqn{eq7}, the forward cross-section, $\sigma^{}_{12}$, has replaced the endothermic one, $\sigma^{}_{3 \gamma}$, using a detailed balance relation between the forward and reverse cross-sections:

	\begin{equation}
		\sigma^{}_{3 \gamma} (E^{}_\gamma) = \frac{g^{}_1 g^{}_2}{g^{}_3(1+\delta^{}_{12})}\, \frac{m^{}_{12} \,c^2 E}{E_\gamma^2} \,\sigma^{}_{12} (E)~,
		\label{eq8}
	\end{equation}
with spin degeneracy factor $g^{}_i = 2J^{}_i +1$, spin $J^{}_i$ and photon energy $E^{}_\gamma =E+Q$. The photon number density, $N^{}_{\gamma q}$ in \eqn{eq7} is also generalised with $q$~\cite{Tirnakl:1997}:
	\begin{equation}
		N^{}_{\gamma q} = \frac{1}{\pi^2 \hbar^3 c^3} \int_0^{\infty} \frac{E_\gamma^2}{\left[1-(1-q)\frac{E^{}_\gamma}{k^{}_B T} \right]^{\frac{1}{q-1}}-1} \,\ud E^{}_\gamma~.
		\label{eq9}
	\end{equation}
It is evident that, for $q\to1$, Eqs.\,\eqref{eq7} and \eqref{eq9} become consistent with the Planck law and Ref.\,\cite{Mathews:2010aa}. Enhancement of the photo-disintegration reaction rate at low temperature, as shown in \fig{fig2}, contributes to drastically reducing the final abundances of the light elements.%

The $q$-dependent energy density of photon is given as~\cite{Buyukkilic:2001ea}
	\begin{equation}
		\rho^{}_\gamma = \frac{(k^{}_B T)^4}{(\hbar c)^3}\,\frac{\pi^2}{15}\,\frac{1}{(4-3q)(3-2q)(2-q)}~.
		\label{eq10}
	\end{equation}
For $q>1$ of concern, photons appear to have additional energy density in comparison with the standard case of $q = 1$. The energy conservation condition at the moment of transition, $\rho^{}_{\gamma\,(q=1)} = \rho^{}_{\gamma\,(q=q')}$ leads to the sudden temperature drop. This temperature drop causes the slight increase in the final abundance of deuterium because it advances the freeze-out time of the light elements due to the reduced two-body nuclear reaction rates. Because we assume that electrons and positrons maintain Fermi Dirac spectra, the drop in temperature causes a loss of energy for the charged leptons and local energy conservation is violated. However, the energy loss of non-relativistic particles due to the temperature drop is sufficiently small in the transition region because $\mathit{\Delta} \rho^{}_e/\rho^{}_e \sim m^{}_e\,\mathit{\Delta} T/T^2 \ll 1$ for small $\mathit{\Delta} T$.

%=====================
\section{Results}
\label{s03}
%=====================
	\begin{figure}[t]
	\centering
		\includegraphics[width=8.5cm]{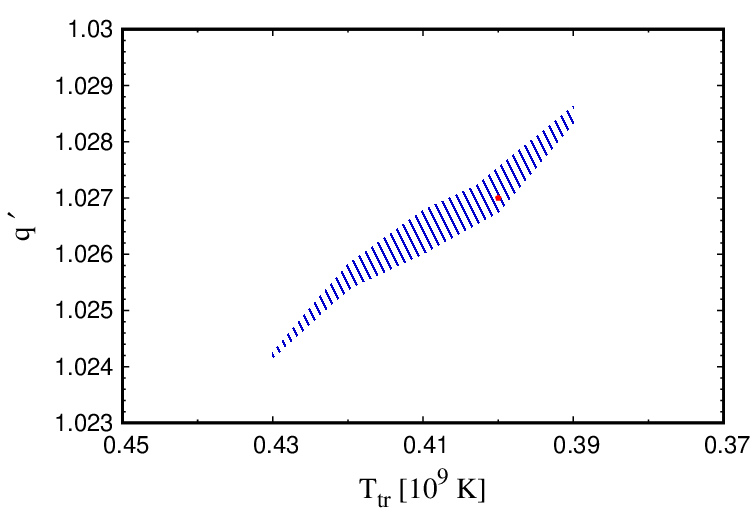}
		\caption{Possible solutions (indicated by the shaded area) in the parameter space. An example ($q\,' = 1.027$, $T^{}_\text{tr} = 4\times10^8\unit{K}$) indicated by the red dot is used in the text to present the resulting abundances.}
		\label{fig3}
	\end{figure}

\begin{figure}[t]
	\centering
		\includegraphics[width=8.5cm]{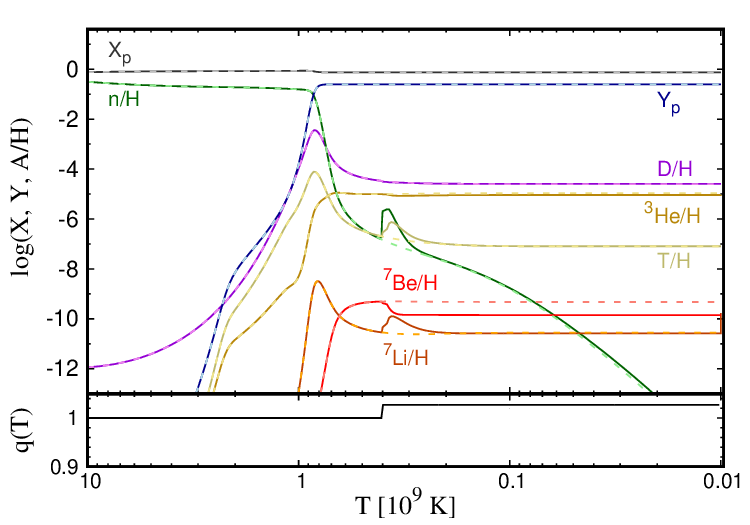}
		\caption{Evolution of light element abundances (upper panel) and $q(T)$ (lower panel) with the values in \eqn{eq11} as a function of temperature. The solid and dashed lines stand for our result and the SBBN, respectively.}
		\label{fig4}
	\end{figure}

Taking all the contributions mentioned so far into account in the BBN network calculation, we found the possible solutions of ($q\,'$, $T^{}_\text{tr}$) that are depicted in \fig{fig3}. The primordial abundances calculated with these solutions are consistent with the up-to-date observations within the errors. As an example of the solutions, \tabl{tab1} shows the abundances for the case of 
	\begin{equation}
		q\,' = 1.027~, \qquad T^{}_\text{tr} = 4\times10^8\unit{K}~,
		\label{eq11}
	\end{equation}
in comparison with the observational data and also the SBBN prediction. The \nuc{He}{4} mass fraction, $Y_\text{p} = 0.2474$, does not differ from the SBBN prediction because it is decoupled at $T > T^{}_\text{tr}$ (i.e. at early times) and unaffected later by the transition to the weakly non-ideal plasma. Adopting the updated $S$-factor of $\mathrm{D}(\mathrm{p},\,\gamma)\nuc{He}{3}$~\cite{Iliadis:2016vkw,Marcucci:2015yla} in the SBBN results in the deuterium abundance, $\mathrm{D/H} = 2.493\times10^{-5}$, which lies marginally outside the observational error boundaries \cite{Cooke:2017cwo}. However, the final deuterium abundance, $\mathrm{D/H} = 2.525\times10^{-5}$, predicted in our model is slightly larger and gets closer to the central value of the observational data. For the \nuc{He}{3} abundance, the difference between the SBBN result and ours is not large (see also \tabl{tab1}), and the reason is that due to a large $Q$-value ($Q=5.493\unit{MeV}$), the $\nuc{He}{3}(\gamma,\,\mathrm{p})\mathrm{D}$ reaction stops before the temperature cools down to $T^{}_\text{tr}$. However the distorted photon distribution in our model makes this destructive process of \nuc{He}{3} reactivated at $T \leq T^{}_\text{tr}$.  Both the \nuc{He}{3} abundances predicted by SBBN and our model are obtained safely below the observational upper limit of  $(1.1 \pm 0.2)\times 10^{-5}$. Finally the \nuc{Li}{7} abundance, which is the major concern of this study, is predicted as $\nuc{Li}{7}/\mathrm{H} = 1.677\times10^{-10}$ and is in remarkable agreement with the observed value\,\cite{Sbordone:2010zi}. Right after the transition, the enhanced reaction rate of $\mathrm{D}(\gamma,\mathrm{n})\nuc{H}{1}$ produces more neutrons, which increase \nuc{Li}{7}/H through the reaction of $\nuc{Be}{7}(\mathrm{n,\,p})\nuc{Li}{7}$ (see the bumps in the abundance curves right below the transition temperature in \fig{fig4}). However, as time goes by, the reaction $\nuc{Li}{7}(\mathrm{p,\,\alpha})\nuc{He}{4}$ continues to destroy \nuc{Li}{7} until equilibrated. (Although the photo-disintegration of  \nuc{Li}{7} is enhanced after the transition, $\nuc{Li}{7}(\mathrm{p,\,\alpha})\nuc{He}{4}$ still dominates the \nuc{Li}{7} destruction.) Therefore, the significant reduction of the final abundance of \nuc{Li}{7} in our calculation compared with that in the SBBN comes from the further destruction of \nuc{Be}{7}, due to the enhanced rates of $\nuc{Be}{7}(\mathrm{n,\,p})\nuc{Li}{7}$ and $\nuc{Be}{7}(\gamma,\,\alpha)\nuc{He}{3}$ after the transition, which eventually decays to \nuc{Li}{7} via electron capture (see \fig{fig4}).

It is also shown in \fig{fig4} that the transition temperature, $T^{}_\text{tr} = 4\times10^8\unit{K}$, is located near the inflection point of the neutron abundance curve, at which the destruction rate of neutron slows down. This means that neutron capture, which continues to convert a simple electron-proton plasma, that is the initial state of BBN, to a more complex plasma having multi-charged nuclides, occurs rapidly before the transition. Moreover $T^{}_\text{tr}$ is surprisingly consistent with the temperature at which electron chemical potential starts to increase~\cite{Pitrou:2018cgg}, although the solution space for $T^{}_\text{tr}$ in \fig{fig3} appears to be outside of or on the border of the temperature region of plasma non-ideality, which is centred at $T \approx 10^9\unit{K}$ in \fig{fig1}. In our proposed solution, the non-ideal plasma region during the BBN epoch corresponds to an increasing point of the chemical potential rather than to the region obtained by the traditional plasma parameter. Our findings support the idea of our BBN scenario with a transition from ideal to weakly non-ideal plasma self-consistently.

	\begin{figure}[t]
	\centering
		\includegraphics[width=8.5cm]{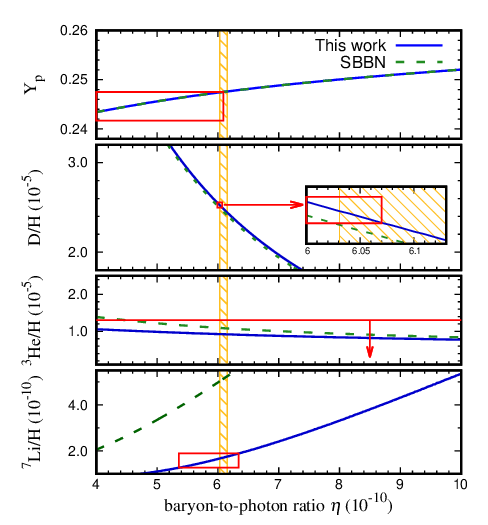}
		\caption{Primordial abundances as a function of baryon-to-photon ratio. Green dashed and blue solid lines indicate the SBBN and our solution, respectively. Red boxes are observational data of each abundance and the shaded yellow area shows the CMB constraint on the baryon-to-photon ratio: $\eta = (6.094 \pm 0.063)\times10^{-10}$. In our model, lithium solutions are found at $6.031 < \eta\times 10^{10} < 6.070$.}
		\label{fig5}
	\end{figure}

Figure\,\ref{fig5} plots the abundance of each element predicted by SBBN and our model as a function of baryon-to-photon ratio. In our model, the baryon-to-photon ratio decreases at the transition due to the distorted photon distribution and we additionally assume that the Planck distribution is restored sometime between the end of BBN and the recombination epoch in order to be consistent with the present CMB observation. From \fig{fig5}, we find that the predicted primordial abundances can constrain the possible value of baryon-to-photon ratio when they are compared to the observations. For example, the strongest constraint on the baryon-to-photon ratio comes from the primordial deuterium abundance, such that the solution to the deuterium abundance cannot be found when $\eta > 6.050\times10^{-10}$. Together with the CMB constraint, we are able to narrow down the possible range of the baryon-to-photon ratio to $6.031 < \eta\times 10^{10} < 6.070$.

%=====================
\section{Restoration of Planck distribution}
\label{s04}
%=====================
Lastly, a question arises as to how the distorted photon spectrum could be restored to the observed blackbody spectrum of CMB whose possible distortions have been limited to $\mathit{\Delta} \rho^{}_\gamma / \rho^{}_\text{CMB}<10^{-5}$ since COBE/FIRAS~\cite{Mather:1993ij,Fixsen:1996nj}. Our model is based upon the idea that the distorted photon distribution is caused by collective effects in the weakly non-ideal plasma. Then, an idea to consider is that the restoration occurs by means of Landau damping defined as the damping of a collective mode of oscillations in a plasma where collisions between the charged particles are negligibly rare; that is, interaction between plasma waves and non-relativistic ions through the plasma epoch between BBN and recombination.

\begin{figure}[t]
	\centering
		\includegraphics[width=8.5cm]{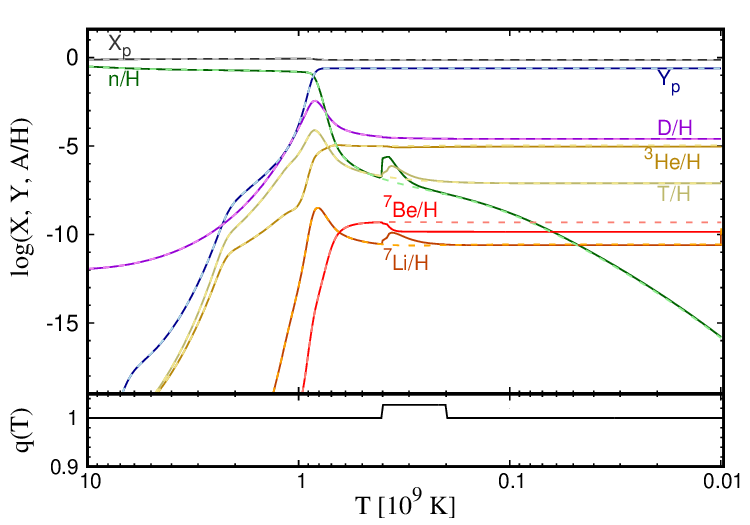}
		\caption{Same as \fig{fig4} but for a transient (box shape) $q(T)$. See text for more details.}
		\label{fig6}
	\end{figure}

In fact, we also found that a transient distortion, such as a box shape, during the BBN epoch could solve the problem if the duration were sufficiently long. This transient distortion would be plausible because in the BBN era any distortion of radiation spectrum is quickly erased through thermalisation processes.  Figure\,\ref{fig6} shows the evolution of light element abundances obtained with the transient $q(T)$ that includes the distortion of $q' = 1.027$ at $T^{}_\text{tr} = 4\times10^8\unit{K}$ and restoration to $q=1$ at $T_\text{re} = 2\times10^8\unit{K}$. This solution predicts the following abundances:
\begin{align}
	Y_\text{p} &= 0.2474~, \nn \\
	\mathrm{D/H} &= 2.503 \times 10^{-5}~, \nn \\
	{}^{3}\mathrm{He/H} &= 0.9322 \times 10^{-5}~, \\
	{}^{7}\mathrm{Li/H} &= 1.664 \times 10^{-10}~, \nn
	\label{eq12}
\end{align}
which also fall within the errors of the observational data. The changes in the photon spectrum at $T<T_\text{re}$ do not affect the primordial abundances of light elements significantly since they are already decoupled. The duration of the distorted photon distribution could be long enough to give the solution to the \nuc{Li}{7} problem, likely during the period that the positron annihilation actively takes place. It can be interpreted that at $T_\text{re}$, right below the temperature at which the annihilation rate has lowered compared to the cosmic expansion rate, the annihilation process can no longer be the source of the chemical disequilibration. Then, Compton processes are able to thermalise the distorted photon spectrum excessively quickly into a blackbody one. Implementing this scenario does not require a modification of the thermal history after BBN. This is an advantage and disadvantage at the same time because there would be no particular measurable quantity to identify this transient distortion hidden behind the blackbody surface.

 Although this is an important question and should be investigated in more detail, further discussion is beyond the scope of the current paper since such non-linear processes are adequately described by kinetic theory.

%=====================
\section{Conclusion}
\label{s05}
%=====================
In this work, we propose a model that could resolve the primordial lithium problem by considering a proper correction to the plasma properties commonly assumed in the standard BBN scenario. In the BBN epoch, thermal equilibrium can be readily attained via Compton-like processes unless there is a source to disturb equilibration. Considering the rapid growth of electron chemical potential, we conjecture that the $e^+e^-$ pair annihilation process could be the source disturbing equilibration that makes the plasma deviate from the ideal state by breaking the chemical equilibrium between the pairs and photons. Perceiving that the BBN plasma could not be perfectly ideal at $T\lesssim 10^9\unit{K}$, we considered the BBN plasma to be weakly non-ideal at $T\leq T^{}_\text{tr}$ and used the distorted photon distribution in order to parametrise the non-ideality that causes the deviation from a blackbody spectrum of the electromagnetic field in a plasma. In the non-ideal plasma, we consider the deviation of the photon spectrum only, more specifically, the increase in the high-frequency regime, which is likely to be caused by the pair annihilation, and its direct consequences such as the reaction rate of photo-disintegration and the photon number density, etc. The perturbative spectral distortion of the photons, whose average energy scale is less than $1\unit{MeV}$, is insufficient to affect the velocity distribution of a nucleus of mass scale $1\unit{GeV}$, even if transferred through the nuclear Compton scattering ($\gamma N \to \gamma\,' N'$). Thus, the nuclear velocity distributions are regarded as Maxwellian. Although there is an issue that the spatial fluctuation of photon density can contribute to the final abundances (especially deuterium abundance, which is sensitive to the changing baryon-to-photon ratio), the plasma in our model is still thought to be homogeneous and isotropic because the Compton processes are efficient enough to erase the temperature gradient and to smooth out the density contrast.

With this correction, the calculated primordial abundances of the light elements are in excellent agreement with the recent observational data up to \nuc{Li}{7} as shown in \tabl{tab1}. There are two main contributions of the distortion of photon distribution to the final abundances. The photo-disintegration processes of light elements are enhanced, which plays a crucial role in reducing the primordial \nuc{Li}{7} abundance. The condition for the energy conservation at the moment of transition, on the other hand, makes the freeze-out time of light elements earlier so that the deuterium abundance is improved. In order to be consistent with the CMB study, we also investigated the possibility that the distorted spectrum of photons can be restored to a blackbody before the BBN era ends. In fact, the relaxation can be allowed at any time since the nuclear abundances are frozen out, also giving a consistent solution. We note that the higher order correction, such as the temperature gradient due to the anisotropy in the radiation field, can cause tension with observations for deuterium that looks more sensitive to $\eta$ and has a small observational error bar. However, this issue applies to all the BBN models, including ours and SBBN, which usually neglect the higher order corrections.

In this work, we propose a model for the non-ideal state of the BBN plasma by using Tsallis statistics. The results based on our model give us a hint that the primordial lithium problem may be solved by the study of plasma characteristics, which will be discussed in more detail in the future studies. To make the study as thorough as possible, we plan to construct and solve the Boltzmann equation for photons under the BBN environment where the pair annihilation must be included properly, as well as the Compton scattering. The exact distortion in the photon distribution will be obtained through these two competing processes.

%=====================
%\begin{acknowledgements}
\section*{Acknowledgements}
%=====================
We are grateful to M.~Kusakabe for useful discussions and information. The work of Y.K. is supported by the National Research Foundation of Korea (NRF) under grant number, NRF-2018R1D1A1B07051239, and also NRF grant funded by the Korea government (MSIT)  (No. 2018R1A5A1025563). The work of D.J. is supported by Institute for Basic Science under IBS-R012-D1. The work of K.K. is supported by NRF-2016R1A5A1013277 and NRF-2016R1D1A1B03936169. The work of M.K.C. is supported by NRF-2020R1A2C3006177 and NRF-2013M7A1A1075764.
%\end{acknowledgements}

%--------------------------------------------------------------------------------------------------------------------------
%       LIst of References
%--------------------------------------------------------------------------------------------------------------------------

\end{document}